\documentclass[preprint]{aastex}
\tighten

\input psfig

\def\kmsm{\,{\rm km\, s^{-1}\, Mpc^{-1}} }

\def\kpc{\,{\rm kpc} }

\begin{document}

\title{Self-Similar Models for the Mass Profiles of Early-type
Lens Galaxies}

\author{D. Rusin\altaffilmark{1}, C.S. Kochanek\altaffilmark{1},
C.R. Keeton\altaffilmark{2,3}}

\altaffiltext{1}{Harvard-Smithsonian Center for Astrophysics, 60
Garden St., Cambridge, MA 02138}

\altaffiltext{2}{Astronomy and Astrophysics Department, University of
Chicago, 5640 S. Ellis Ave., Chicago, IL 60637}

\altaffiltext{3}{Hubble Fellow}

\begin{abstract}

We introduce a self-similar mass model for early-type galaxies, and
constrain it using the aperture mass-radius relations determined from
the geometries of 22 gravitational lenses. The model consists of two
components: a concentrated component which traces the light
distribution, and a more extended power-law component ($\rho \propto
r^{-n}$) which represents the dark matter. We find that lens galaxies
have total mass profiles which are nearly isothermal, or slightly
steeper, on the several-kiloparsec radial scale spanned by the lensed
images. In the limit of a single-component, power-law radial profile,
the model implies $n=2.07\pm 0.13$, consistent with isothermal
($n=2$). Models in which mass traces light are excluded at $>99\%$
confidence. An $n=1$ cusp (such as the Navarro-Frenk-White profile)
requires a projected dark matter mass fraction of $f_{cdm} =
0.22\pm0.10$ inside 2 effective radii. These are the best statistical
constraints yet obtained on the mass profiles of lenses, and provide
clear evidence for a small but non-zero dark matter mass fraction in
the inner regions of early-type galaxies. In addition, we derive the
first strong lensing constraint on the relation between stellar
mass-to-light ratio $\Upsilon$ and galaxy luminosity $L$: $\Upsilon
\propto L^{0.14^{+0.16}_{-0.12}}$, which is consistent with the
relation suggested by the fundamental plane. Finally, we apply our
self-similar mass models to current problems regarding the
interpretation of time delays and flux ratio anomalies in
gravitational lens systems.

\end{abstract}

\keywords{galaxies: elliptical and lenticular -- galaxies: structure
-- gravitational lensing}

\section{Introduction}

The structure of galaxies is closely linked to their formation and
evolution, and therefore provides a vital testing ground for the cold
dark matter (CDM) paradigm which has proven so successful on large
scales (e.g., Spergel et al.\ 2003). The primary observable is the
shape of the radial mass profile, from which the relative
distributions of luminous and dark matter can be determined. The
outermost regions of galaxies have now been extensively probed using
weak lensing (e.g., Fischer et al.\ 2000; McKay et al.\ 2001;
Kleinheinrich et al.\ 2003) and satellite dynamics (e.g., Zaritsky et
al.\ 1997; Romanowsky \& Kochanek 2001; McKay et al.\ 2002; Prada et
al.\ 2003), each of which reveals an extended dark matter halo out
to $100-300 \kpc$.  Methods for tracing the inner regions of galaxies
largely depend on the galaxy morphology. Spiral galaxies are the most
easily studied, and are consequently the best understood, as rotation
curves can be mapped to $\sim 30 \kpc$ using kinematic tracers such as
HI. Flat rotation curves at $\ga 5 \kpc$ clearly indicate the presence
of dark matter (e.g., Rubin, Thonnard \& Ford 1980), but dynamical
observations also suggest that the inner halo may be significantly
less concentrated (e.g., McGaugh \& de Blok 1998; de Blok et al.\
2001; Salucci 2001; de Blok \& Bosma 2002) than the cuspy halos
predicted by CDM simulations (e.g., Moore et al.\ 1999b; Bullock et
al.\ 2001).

Compared to their late-type counterparts, early-type (elliptical and
lenticular; E/S0) galaxies are not as well understood.  Some of the
best evidence for dark matter outside of a few optical radii has been
obtained from investigations of the extended, X-ray emitting gas
(e.g., Fabbiano 1989; Matsushita et al.\ 1998; Loewenstein \& White
1999). The inner regions of early-type galaxies are typically probed
using stellar dynamics (e.g., Rix et al.\ 1997; Gerhard et al.\
2001). However, the velocity dispersion is not an unambiguous tracer
of the mass profile, as it requires an understanding of the orbital
anisotropy. In addition, the existence of a tight fundamental plane
(FP; Djorgovski \& Davis 1987; Dressler et al.\ 1987) presumably
provides clues about the inner structure of early-type galaxies, but
little consensus has been achieved regarding its meaning (e.g., Faber
et al.\ 1987; Renzini \& Ciotti 1993; van Albada, Bertin \& Stiavelli
1995; Ciotti, Lanzoni \& Renzini 1996; Pahre, de Carvalho, \&
Djorgovski 1998; Bertin, Ciotti, \& Del Principe 2002). Many
outstanding questions therefore remain regarding the structure of E/S0
galaxies, the universality of this structure, and its relation to the
predictions of CDM models.

Strong gravitational lensing is a powerful and unique tool for
investigating the mass distributions in early-type galaxies at
intermediate redshift, as it probes mass directly. First,
model-independent projected masses may be inferred from the geometry
of the lensed images (e.g., Schneider, Ehlers \& Falco 1992). When
mass models are normalized by this constraint, the stellar velocity
dispersion becomes a far more sensitive probe of the radial mass
profile. Ongoing efforts by the Lens Structure and Dynamics (LSD)
survey (Koopmans \& Treu 2002, 2003; Treu \& Koopmans 2002a;
collectively ``KT'') have made significant progress in obtaining and
interpreting dynamical measurements in gravitational lens systems.
Second, models of the galaxy mass distribution can be constructed
based on the positions, morphologies and flux ratios of the lensed
images. Of particular importance are the additional astrometric
constraints obtained from complex structure in the lensed source,
which can be used to break degeneracies and distinguish among radial
mass profiles. These constraints include large-scale extended emission
from radio lobes (e.g., Kochanek 1995) or quasar host galaxies (see,
e.g., Kochanek, Keeton, \& McLeod 2001 and Saha \& Williams 2001 for
general discussions), lensed images of multi-component sources (e.g.,
Cohn et al.\ 2001; Mu\~noz, Kochanek \& Keeton 2001), and the relative
orientations of milliarcsecond-scale radio jets (e.g., Rusin et al.\
2002; Winn, Rusin \& Kochanek 2003).

The mass distributions in several lens galaxies have now been
investigated using either direct modeling or stellar dynamics. In all
but one case, the profile is consistent with an isothermal model
($\rho \propto r^{-2}$). The lone exception is PG1115+080, for which
the high stellar velocity dispersion (Tonry 1998) suggests a mass
profile steeper than isothermal ($\rho \propto r^{-2.35}$; Treu \&
Koopmans 2002b), although this also implies that the galaxy lies well
off the FP. The lensing results are therefore generally consistent
with the nearly isothermal profiles favored by X-ray and dynamical
studies of local elliptical galaxies.

Understanding the mass profiles of lens galaxies is required for the
determination of the Hubble constant ($H_0 \equiv 100h\kmsm$) from
measured time delays. Kochanek (2002) demonstrates that the predicted
delay between two images, and hence the derived value of $H_0$, is
primarily governed by the average surface mass density in the annulus
defined by their radial distances from the lens center, with a small
correction for the slope of the profile in this annulus. Each of these
quantities is determined by the shape of the radial mass profile.
Based on a study of several lens systems which are dominated by a
single galaxy and have well-measured time delays, Kochanek (2002)
finds $h = 0.51\pm 0.05$ if lenses are isothermal. This value is lower
than those obtained from the Hubble Space Telescope (HST) Key Project
($h=0.72 \pm 0.08$; Freedman et al.\ 2001) and WMAP data ($h=0.72\pm
0.05$, although this requires the assumption of a flat cosmology and a
dark energy equation of state of $w=-1$; Spergel et al.\ 2003). In
fact, Kochanek (2002, 2003) shows that early-type galaxies must be
described by models in which mass traces light, and hence have no
extended dark matter halos, to reconcile current time delay
measurements with $h\simeq 0.7$ (see also Williams \& Saha 2000). This
apparent discrepancy with other observational constraints and the CDM
paradigm demands continued examination of the lens galaxy population.

Decomposing the luminous and dark matter components of early-type
galaxies is also potentially important for investigating the nature of
flux ratio discrepancies in gravitational lens systems. It has now
been extensively demonstrated that simple, smooth mass models have
great difficulty reproducing the observed flux ratios in four-image
lenses (e.g., Dalal \& Kochanek 2002; Metcalf \& Zhao 2002; Chiba
2002; Keeton, Gaudi \& Petters 2002), both in the optical and
radio. The effect has been traced to small-scale structure in the
gravitational potential (Kochanek \& Dalal 2003), and can be due to
either stars (microlensing; e.g., Witt, Mao, \& Schechter 1995;
Schechter \& Wambsganss 2002) or CDM satellite halos (millilensing;
e.g., Mao \& Schneider 1998; Metcalf \& Madau 2001; Bradac et al.\
2002; Chiba 2002; Dalal \& Kochanek 2002). Both phenomena are surely
present, as radio sources are generally too large to be affected by
stars, while uncorrelated time variability is an unmistakable
signature of stellar microlensing in some optical lenses (e.g.,
Wozniak et al.\ 2000). Microlensing is most efficient at accounting
for observed anomalies, in particular the preferred demagnification of
the saddle point image, when stars comprise a small fraction
(15--30\%) of the surface mass density at the image radius (Schechter
\& Wambsganss 2002). If we can constrain a two-component model of the
galaxy mass distribution, then we can predict the stellar surface mass
fraction at the radii typically spanned by lensed images. Thus, the
predictions and requirements of the microlensing hypothesis may be
tested.

Because mass distributions have been directly constrained in only a
small number of gravitational lenses, it is worthwhile to consider
statistical constraints on the lens galaxy population. Such attempts
have thus far been very limited.  Several analyses (e.g., Rusin \& Ma
2001; Evans \& Hunter 2002) have focused on the absence of central
images in deep radio maps of gravitational lens systems, which has
been invoked to rule out profiles that are much shallower than
isothermal or have significant cores. Keeton (2003) suggests that such
constraints have been over-interpreted, since the magnification of the
missing image is strongly dependent on the properties of the mass
profile very close to the galaxy center. In addition, because mass
distributions steeper than isothermal do not produce central images,
any constraints based on the absence of these images are
one-sided. The number and sizes of lens systems have also been used to
investigate galaxy mass profiles in the context of CDM models (Keeton
2001). It is shown that adiabatic compression (e.g., Blumenthal et
al.\ 1986) yields nearly flat rotation curves for a range of initial
halo concentrations and cooled baryon fractions (see also Kochanek \&
White 2001 and Kochanek 2003), but typical CDM halos (e.g., Bullock et
al.\ 2001) appear to be too centrally concentrated to account for the
data. The analysis, however, offers few quantitative results
describing the mean mass profile and scatter, and therefore demands
further study of two-component mass models.

In this paper we introduce a new technique to statistically constrain
the mass profiles of early-type lens galaxies via aperture mass
measurements. Each lens must obey a strict relationship between the
image radii and the projected masses they enclose. The set of
mass-radius relations can be used to investigate the typical mass
profile in a sample of lenses if we assume that early-type galaxies
have a self-similar (homologous) structure. This means that the
functional form (shape) of the mass distribution is the same from
galaxy to galaxy, and is scaled by the properties of the light
distribution. An underlying homology between mass and light is
suggested by the existence of a tight fundamental plane relating
optical and dynamical observables. In one popular interpretation of
the FP (e.g., Faber et al.\ 1987; van Albada et al.\ 1995), early-type
galaxies are structurally homologous, but the mass-to-light ratio has
a luminosity dependence. Gerhard et al.\ (2001) use dynamical data to
bolster this hypothesis, demonstrating that more luminous early-type
galaxies have higher stellar mass-to-light ratios in their central
regions. Other analyses have cast doubt on the description of the FP
in terms of strong homology and varying mass-to-light ratios (e.g.,
Caon, Capaccioli, \& D'Onofrio 1993; Pahre et al.\ 1998; Bertin et
al.\ 1994, 2002), and instead suggest a weak homology in which
structural parameters vary systematically with luminosity. However,
because a strong homology is the simplest reasonable assumption, it is
a good starting point for the statistical investigation of mass
profiles in early-type galaxies.

Section 2 details the aperture mass-radius constraints offered by
lensing. Section 3 describes the lens sample and outlines our various
assumptions. Section 4 introduces our self-similar models and the
associated methodology. Section 5 presents our statistical constraints
on the radial mass distribution in lenses. Section 6 demonstrates how
these constraints can be used for the interpretation of measured time
delays and flux ratio anomalies among lensed images. Section 7
discusses our findings in the context of recent astrophysical and
cosmological results. Unless otherwise stated, we assume a flat
$\Omega_M = 0.3$ cosmology with $h=0.65$ for all calculations. 

\section{Aperture Masses from Lens Geometries}

The geometries of gravitational lens systems yield model-independent
constraints on projected masses. Consider, for example, a deflector
with a circularly symmetric surface density distribution, and a source
which sits directly behind it. The image will form on a ring of
(physical) Einstein radius $R_{Ein}$, which is related to the
aperture mass ($M_{Ein}$) it encloses by 
\begin{equation}
\frac{1}{\pi} \frac{M_{Ein}}{R_{Ein}^2} = \frac{c^2}{4\pi G}
\frac{D_s}{D_d D_{ds}} \equiv \Sigma_{cr} \ ,
\end{equation} 
where $\Sigma_{cr}$ is the critical surface density.  Angular diameter
distances to the lens, to the source, and from the lens to the source
are $D_d$, $D_s$ and $D_{ds}$, respectively. For this simple case, the
mass enclosed by the ring is determined by its radius alone, and is
therefore independent of the shape of the mass profile.

Real lenses are more complicated. First, sources (and their images)
are, in general, not symmetrically placed with respect to the mass
distribution. This means that the geometry may be described by more
than one characteristic radius. Second, gravitational potentials are,
in general, not circularly symmetric. The presence of a quadrupole,
due to either external shear or ellipticity, effectively stretches the
image plane along one axis and compresses it along a perpendicular
axis, thereby confusing the interpretation of geometry in terms of
mass. These issues are surmountable, however, allowing any lens
geometry to yield model-independent relations between masses and
radii.

Four-image lenses (quads) offer the best mass constraints, as the
above complications are minimized. The images typically reside at
similar radii, and their good angular coverage allows for the robust
determination and removal of the quadrupole. Just about any mass
profile can reproduce the image positions and fluxes in a quad, but
modeling demonstrates that there is a radius at which all models agree
on the enclosed projected mass (e.g., Cohn et al.\ 2001; Mu\~noz et
al.\ 2001). We have tested the generality of this claim using Monte
Carlo simulations. First, we create fake four-image lenses from an
elliptical mass distribution with some fixed radial profile residing
in an external shear field of random orientation. We then model the
image positions and fluxes using a broad range of mass profiles,
tabulating the mass enclosed by a circular aperture as a function of
radius for each best-fit model. We find that the suite of aperture
mass curves $M(R)$ cross at nearly the same radius ($R_{Ein}$), the
value of which can be accurately estimated (to $\sim 1\%$) from the
critical radius of a model-fit singular isothermal sphere (SIS) in an
external shear field. This is true regardless of whether the mass
distribution is spherical or elliptical. The enclosed, projected mass
inside $R_{Ein}$ is then given by eq.~(1), to an accuracy of $\sim
2\%$.

Two-image lenses (doubles) yield weaker mass constraints. If the
potential is circularly symmetric, then the lens equation dictates a
model-independent mass-radius relation:
\begin{equation}
\frac{1}{\pi} \left[ \frac{M(R_1)}{R_1} + \frac{M(R_2)}{R_2}
\right] \left( \frac{1}{R_1 + R_2} \right) =  \Sigma_{cr} ,
\end{equation}
where $R_1$ and $R_2$ are the radii of the images with respect to the
galaxy center, and $M(R_1)$ and $M(R_2)$ are the projected masses they
enclose. Hence, doubles constrain a combination of two radii and two
masses. Note that in the limit of a ring ($R_1 = R_2$), we recover
eq.~(1). The presence of a quadrupole smears the mass-radius relation
for doubles, and the limited number of constraints means that its
effects cannot be removed in a model-independent manner. If the
quadrupole is represented as an external shear, then the maximum
fractional deviation from eq.~(2) is equal to the shear magnitude
$\gamma$, and this occurs when the shear axis is parallel or
perpendicular to the axis defined by the lensed images. Averaging over
orientations, the rms deviation is $\simeq \gamma/\sqrt{2}$. We have
confirmed this assertion using Monte Carlo simulations, in which fake
doubles are produced from one mass profile and modeled using a range
of mass profiles. While four-image lenses typically have quadrupoles
which can be described by shear fields of amplitude $0.1 < \gamma
<0.2$ (e.g., Keeton, Kochanek \& Seljak 1997; Holder \& Schechter
2003), we might expect doubles to have somewhat lower quadrupoles, on
average, as they represent a greater fraction of the lensing cross
section when the shear is small (e.g., Keeton et al.\ 1997; Rusin \&
Tegmark 2001; Finch et al.\ 2002). Consequently, deviations from
eq.~(2) should generally be $\sim 10\%$. We therefore employ eq.~(2)
as a model-independent mass-radius relation for doubles, and set a
$10\%$ tolerance for quadrupole effects.\footnote{We note in passing
that when using steep radial profiles to fit fake lenses produced from
shallower profiles, huge external shears or extremely flattened
deflectors may be required. These solutions, however, can be rejected
on the grounds of physical implausibility. First, the mass
distributions of lens galaxies would have to be significantly more
flattened than the light distributions, which typically have moderate
axial ratios of $\sim 0.7-0.8$; (e.g., Keeton et al.\ 1997). Second,
such large external shears would have to be directly traced to nearby,
cluster-scale perturbers, which are not observed in the relatively
low-density environments of lens galaxies (Keeton, Christlein \&
Zabludoff 2000).}

\section{Data}

\subsection{Lens Sample}

We begin with the sample of 28 early-type gravitational lens galaxies
and bulges analyzed by Rusin et al.\ (2003). Note that this includes
Q2237+050 (Huchra et al.\ 1985), which is lensed entirely by the bulge
of a spiral galaxy. Twenty-two of the systems have measured lens
redshifts ($z_d$), and 18 of these are in combination with a measured
source redshift ($z_s$).\footnote{Here we include a recent measurement
of $z_s = 1.17$ (Treu \& Koopmans 2003) for MG1549+3047 (Leh\'ar et
al.\ 1993).}  Previously, for systems with no $z_d$, the lens redshift
had been estimated under the requirement that the galaxy fall on the
fundamental plane (Kochanek et al.\ 2000). To make use of this
technique, one must first estimate the stellar velocity dispersion,
which necessitates an explicit assumption of a galaxy mass
profile. Since this paper is concerned with constraining mass models,
it is prudent to exclude the six lenses without spectroscopic lens
redshifts. The number of lenses used in the following analysis is
therefore 22. The systems are listed in Table~1.

\subsection{Geometric Properties}

The important geometric properties of each lens system are derived
from the radii of the lensed images. Image positions with respect to
the lens galaxy are determined from the HST data, based on the fitting
methods outlined by Leh\'ar et al.\ (2000).\footnote{ Because the two
images of the radio lens B2319+051 (Rusin et al.\ 2001) have not been
detected in the optical, a direct determination of their positions
with respect to the galaxy is currently impossible. However, the
structure in the lensed jets allows the galaxy position to be
determined from mass modeling. Rusin et al.\ (2001) demonstrate this
in the context of an isothermal mass profile, but our calculations
show that the recovered galaxy position is nearly independent of the
assumed model. We therefore include this system in our analysis, using
the estimated $R_1$ and $R_2$ from that paper.}  The uncertainties on
these positions are typically $\la 5$ milliarcseconds, which is
negligible compared to the radii of even the closest lensed images in
doubles ($\sim 0.1$ arcsec). For quads and rings, the angular Einstein
radius $r_{Ein}$ is determined by fitting an SIS plus external shear
model to the lens data, and then converted to the physical radius
$R_{Ein} = r_{Ein} D_{d}$. The Einstein radius can typically be
derived to $\sim$1\% precision, so its uncertainty is also
negligible. The geometric properties of the lens sample are listed in
Table~1.

Interpreting radii in terms of masses requires the critical surface
density $\Sigma_{cr}$, which depends on source and galaxy redshifts
through the angular diameter distances. Recall that each lens galaxy
in our sample has a measured redshift. If the source redshift has also
been measured, then there is no uncertainty in $\Sigma_{cr}$, at least
within the context of our standard cosmological assumptions. If the
source redshift is not known, we derive the uncertainty in
$\Sigma_{cr}$ (actually $\log \Sigma_{cr}$) with Monte Carlo
techniques. Following Rusin et al.\ (2003), we draw 10000 values from
a Gaussian distribution $z_s = 2.0 \pm 1.0$ for the source redshift,
keeping only trials with $1 \leq z_s \leq 5$ because virtually all
known lensed sources lie within this range. We adopt the $z_s = 2.0$
value of $\log \Sigma_{cr}$ as the median, and the rms scatter around
this value as the uncertainty in $\log \Sigma_{cr}$.

\subsection{Photometric Properties}

The important photometric properties of each galaxy are the
intermediate axis effective radius and luminosity. These quantities
are derived from HST data obtained with the WFPC2 and NICMOS
cameras. The fitting methods are detailed by Leh\'ar et al.\ (2000),
Kochanek et al.\ (2000) and Rusin et al.\ (2003). The optical
effective radius ($r_e$; or $R_e = r_e D_d$ in physical units) is
determined by fitting a de Vaucouleurs profile to the galaxy in the
filter with the highest signal-to-noise ratio.  This quantity is then
held fixed to determine the mean surface brightness within the
effective radius ($\mu_{e,Y}$) in all filters $Y$. The total magnitude
is then $m_Y = \mu_{e,Y} - 5 \log r_e - 2.5 \log 2\pi$. The
photometric properties of our sample are tabulated in Rusin et al.\
(2003).

Complications arise from the need to convert the observed galaxy
magnitudes to some standard luminosity scale that accounts for
$K$-corrections and luminosity evolution.  As we shall see in \S 4,
only the magnitude offset $M-M_* = -2.5 \log (L/L_*)$ is required for
our analysis, since we can parameterize the homology model in terms of
the present-day ($z=0$) stellar mass-to-light ratio for an $L_*$
galaxy. Assuming that $M-M_*$ does not evolve (i.e., that the
evolution rate is independent of luminosity; the same assumption was
made in Rusin et al.\ 2003 and all other analyses using the FP to
study galaxy evolution; e.g., van Dokkum et al.\ 1998, 2001; Treu et
al.\ 2001, 2002), we can estimate this quantity from the observed
magnitudes, without converting to rest frame bands.  Specifically, we
compare the observed magnitudes ($m_{obs,Y}$) to models ($m_{mod,Y}$)
for the apparent magnitudes of an $L_*$ galaxy at the lens redshift,
and then calculate the mean offset:
\begin{equation}
\langle M-M_* \rangle = \frac{\sum_Y[m_{obs,Y}-m_{mod,Y}]/(\delta
m_{obs,Y})^2}{\sum_Y 1/(\delta m_{obs,Y})^2} \ .
\end{equation}
We compute the model magnitudes by convolving spectral energy
distributions from the GISSEL96 version of the Bruzual \& Charlot
(1993) spectral evolution models with transmission curves for HST
filters (available from the technical archives of STScI, with
zero-points from Holtzman et al.\ 1995).  All models are normalized to
a fixed present-day characteristic magnitude of $M_{*} = -19.9 + 5
\log h$ for early-type galaxies in the $B$ band (e.g., Madgwick et
al.\ 2002). We take as a fiducial model an instantaneous starburst at
$z_f=3$ with solar metallicity $Z=Z_{\odot}$ and a Salpeter (1955)
initial mass function (IMF).  The assumed metallicity is consistent
with observations of early-type field galaxies at $0.3 < z < 0.9$
(Ferreras, Charlot \& Silk 1999), and the assumed formation redshift
is consistent with a number of analyses which strongly favor old
stellar populations (mean star formation redshift $\langle z_f\rangle
> 2$; e.g., Bernardi et al.\ 1998; Schade et al.\ 1999; Kochanek et
al.\ 2000; van Dokkum et al.\ 2001; Im et al.\ 2002; van de Ven, van
Dokkum \& Franx 2002; Rusin et al.\ 2003). Considering a broad range
of stellar models ($1.5 < z_f < 5$, $0.4 < Z/Z_{\odot} < 2.5$) yields
an rms scatter in $\langle M-M_*\rangle$ of 0.1--0.2~mag for most
lenses. This uncertainty is much larger than the measurement errors,
since all of the 22 galaxies have excellent photometry.  We therefore
adopt a luminosity error of $\delta (M-M_*) = 0.20$, or $\delta \log
(L/L_*) = 0.08$, for all lens galaxies.

Estimating $L/L_*$ requires some understanding of luminosity
evolution, which is implicitly included in the spectral models. For
example, when normalized to a fixed $L_{*}$ at $z=0$, a faster
evolution rate (lower $\langle z_f \rangle$) would predict higher
values of $L_*(z)$, and hence we would derive lower values of $L/L_*$
for each lens galaxy. If the true mean star formation redshift is
$\langle z_f \rangle < 1.5$, then our fiducial model of $\langle z_f
\rangle = 3$ could greatly overestimate $L/L_*$, especially for high
redshift lenses ($z_d > 0.7$). However, most studies of early-type
field galaxies are inconsistent with the rapid luminosity evolution
implied by $\langle z_f \rangle < 1.5$.  We note in passing that if
the evolution rate is explicitly fit by our mass models (\S 4), the
scatter is minimized by $\langle z_f \rangle \simeq 3$, and increases
for later formation redshifts. Unfortunately, the resulting
constraints on the evolution rate are very poor, as a number of other
parameters are being fit. Since this paper focuses on the structure of
lens galaxies, we believe it is best to simply assume a reasonable
mean star formation redshift of $\langle z_f \rangle=3$. We reiterate
that our estimated uncertainties on $L/L_*$ are sufficient to account
for the spread in evolution rates over the favored range of $\langle
z_f \rangle > 1.5$.

Finally, note that many of the quantities entering our analysis are
correlated.  For example, the total magnitude and the fraction of the
light enclosed by the aperture radius $R$ both correlate with the
effective radius $R_e$.  Neither effect is very significant, because
$\delta \log R_e$ is typically quite small (see Rusin et al.\ 2003),
but for completeness we take these correlations into account via Monte
Carlo calculations.  We draw 10000 trials from a Gaussian distribution
representing the effective radius (mean $\log R_e$, width $\delta \log
R_e$), compute the derived quantities, and store them for later use.
We use the 10000 sets of correlated parameters, plus the independent
(uncorrelated) luminosity error of $\delta \log (L/L_*) = 0.08$, to
calculate the scatter in the quantities entering the fitting functions
given below.

\section{Self-similar Models for Galaxy Mass Distributions}

The aperture mass-radius relation (eq. 1 or 2) for a single lens tells
us nothing about the mass profile. However, recall that the image
splitting scale depends on the redshifts of both the lens galaxy
($z_d$) and lensed source ($z_s$) through the angular diameter
distances in $\Sigma_{cr}$. For example, consider a spherical galaxy
with a power-law mass density $\rho(r) = \rho_0 (r/r_0)^{-n}$, and a
corresponding surface density $\Sigma(R) = \Sigma_0 (R/R_0)^{1-n}$. If
a source sits directly behind this galaxy, it will be lensed into a
ring of Einstein radius $R_{Ein} = [2 \Sigma_0 / \Sigma_{cr} (3-n)
]^{1/(n-1)} R_0$. We could therefore imagine placing a fixed mass
distribution at various $z_d$, where it will lens sources at various
$z_s$. The resulting set of Einstein radii $R_{Ein}$ and aperture
masses $M_{Ein}$ would then allow us to trace out the radial mass
profile of that galaxy.

Nature offers us an ensemble of gravitational lens systems with a wide
range of lens ($0 \la z_d \la 1$) and source ($1 \la z_s \la 5$)
redshifts, but the galaxy properties are far from uniform. The
aperture mass-radius relations can, however, determine the typical (or
``mean'') mass profile of the lens galaxy population if all galaxies
can be placed on a common mass scale. This is most easily accomplished
by postulating some self-similar (homology) model to relate the mass
and light distributions of early-type galaxies. A given model allows
projected masses $M(R)$ to be predicted for each lens using the
photometric parameters, which can then be compared to the mass-radius
relations determined from the lensing geometries. The large range of
image radii in our sample, spanning $0.2 \la R/R_e \la 7$, gives us a
long baseline for mapping the profile. We note, however, that the
distribution is not uniformly populated, with most of the quads and
rings probing the range $1 \la R_{Ein}/R_e \la 4$. Doubles help fill
in the radial coverage: the inner images typically span $0.4 \la
R_1/R_e < 1$, while the outer images span $2 \la R_2/R_e \la 4$.

We consider a physically-motivated two component (luminous plus dark
matter) model for the mass distribution. Because of the limited sample
of lenses on which to test the model, we keep the number of parameters
to a minimum. The component which traces the luminous matter is
modeled by two parameters: the mean present-day ($z=0$) stellar
mass-to-light ratio $\Upsilon_*$ (in the rest frame $B$ band) for an
$L_*$ galaxy, and an exponent ($x$) describing its dependence on the
galaxy luminosity $L$: $\Upsilon \propto L^x$. Under the assumption of
homology, the fundamental plane implies that $x \simeq 0.3$ (e.g.,
Jorgensen, Franx \& Kjaergaard 1996). For each galaxy, the luminous
surface mass density is represented by a de Vaucouleurs profile, and
the associated mass enclosed by an aperture $R$ is
\begin{equation}
M_{lum}(R) =
\Upsilon_*\,L_*\,\left(\frac{L}{L_*}\right)^{1+x}\,g\left(\frac{R}{R_e}\right)
\ ,
\end{equation}
where $g(R/R_e)$ is the projected fraction of the luminosity inside
$R$.  Recall that for the de Vaucouleurs profile, $g(0) = 0$, $g(1) =
0.5$ and $g(\infty) = 1$. Note that our use of the intermediate axis
effective radius, which is the geometric mean of the major and minor
axes, allows us to accurately determine the fraction of the luminosity
enclosed by a circular aperture, even when the surface brightness
distribution is elliptical.

We model the dark matter as a power-law mass distribution. N-body
simulations predict halos in which the mass density follows a shallow
power-law slope at small radius, and turns over to a steeper slope
beyond some break radius (e.g., Navarro, Frenk \& White 1997,
hereafter NFW; Moore et al.\ 1999b). However, because the lensed image
radii probe only the inner several kiloparsecs of the mass
distribution, we believe that a power-law approximation for the dark
matter halo is reasonable on this scale. We describe this component by
two parameters: the projected mass fraction ($f_{cdm}$) within $2
R_e$, which is roughly the median scale of the Einstein radii of our
lens sample, and the logarithmic slope of the mass density profile
($n$, where $\rho \propto r^{-n}$).\footnote{The parameter $n$ should
be thought of as a ``local'' density slope. In reality, our fits use a
surface density $\Sigma \propto R^{1-n}$. This avoids the divergences
encountered in converting between volume and surface densities for
$n\leq 1$.} For each galaxy, the dark matter enclosed by an aperture
$R$ is
\begin{equation}
M_{cdm}(R) = M_{lum}(2 R_e) \ \frac{f_{cdm}}{1-f_{cdm}} \
 \left(\frac{R}{2 R_e}\right)^{3-n} = \Upsilon_*\,L_*
 \left(\frac{L}{L_*}\right)^{1+x}\, g(2)\ \frac{f_{cdm}}{1-f_{cdm}}
 \left(\frac{R}{2 R_e}\right)^{3-n} \ ,
\end{equation}
where $g(2) = 0.69$. 
The total model-predicted mass inside $R$ is then 
\begin{equation}
M_{mod}(R) = M_{lum}(R) + M_{cdm}(R) \ .
\end{equation}
Note that the parameters $\Upsilon_*$ and $x$ set the normalization of
the mass profile, while the dark matter parameters $f_{cdm}$ and $n$
modulate its shape. The homology scheme effectively places all
galaxies on a common mass scale by normalizing the aperture masses
$M(R)$ by $(L/L_*)^{1+x}$, and the aperture radii $R$ by $R_e$. If a
strict homology holds, then all galaxy aperture masses and radii
should reside on a curve in the space of $M(R)/(L/L_*)^{1+x}$ versus
$R/R_e$.

Modeling, dynamical and statistical studies of gravitational lens
galaxies often make use of a single-component mass distribution with a
scale-free radial profile. While simplistic, this model allows one to
explore much of the lensing phenomenology related to mass
concentration, and provides a standard for comparing profile
constraints from different lens systems. It is therefore worthwhile to
consider the scale-free limit of our homology model. By taking
$\Upsilon_* \rightarrow 0$ and $f_{cdm} \rightarrow 1$ (such that
$\Upsilon_* f_{cdm} / (1-f_{cdm}) \rightarrow {\rm const.}$), we
obtain a pure power-law mass distribution:
\begin{equation}
M_{pl}(R) = M_{0} \left(\frac{L}{L_*}\right)^{1+x}
\left(\frac{R}{2 R_{e}}\right)^{3-n} \ ,
\end{equation}
where $M_0$ is the projected mass inside $2 R_e$ for an $L_*$ galaxy.

We will constrain the parameters of the homology model by the set of
mass-radius relations from the lens sample. The optimization is
performed using the goodness-of-fit function
\begin{equation}
\chi^2 = \chi^2_{qr} + \chi^2_{d} \ .
\end{equation}
In quads and rings, the lensing geometry constrains the mass enclosed
by the Einstein radius (eq.~1). These systems are evaluated using
\begin{equation}
\chi^2_{qr} = \sum_i \left[ \frac{ \log M_{mod}(R_{Ein,i})
 - \log [\Sigma_{cr,i} \pi R_{Ein,i}^2 ]}{\delta_{scale} \delta_i} \right]^2
\ \ ,
\end{equation}
where $\delta_{scale}$ and $\delta_i$ are defined below.  In doubles,
the lensing geometry constrains a combination of two radii and
projected masses (eq.~2). These systems are evaluated using
\begin{equation}
\chi^2_{d} = \sum_i \left[ \frac{ \log [M_{mod}(R_{1,i})/R_{1,i} +
M_{mod}(R_{2,i})/R_{2,i}] - \log[\Sigma_{cr,i} \pi (R_{1,i} +
R_{2,i})] }{\delta_{scale} \delta_i} \right]^2 \ .
\end{equation}
The logarithmic uncertainty $\delta_i$ on each data point is derived
using the Monte Carlo methods outlined in \S 3, but is well
approximated as $\delta_i^2 \simeq (1+x)^2 (\delta \log L/L_*)^2 +
(\delta \log \Sigma_{cr,i})^2 + \delta^2_{\gamma}$, where
$\delta_{\gamma}$ is the additional 10\% tolerance for
quadrupole-related smearing of the mass-radius relation in doubles. We
set $\delta_{\gamma} = 0$ for quads and rings, and $\delta \log
\Sigma_{cr,i} = 0$ for systems with a measured $z_s$. Finally, we note
that the above procedures have been tested using Monte Carlo
simulations, and we find that the input mass profiles can be
successfully recovered.

Following optimization, we uniformly rescale the estimated errors by
setting $\delta_{scale}$ so that the best-fit model has $\chi^2 =
N_{DOF}$, the number of degrees of freedom. This does not alter the
optimized parameters, but does allow us to relate the uncertainties to
the observed scatter in the homology model. Moreover, it preserves the
relative weighting among the data points, which naturally gives more
weight to quads and rings, and those systems with a measured
$z_s$. Rescaling is particularly important because our model is
undoubtedly a simplistic representation of galaxy mass distributions.
Hence, the $\chi^2$ is likely to be dominated by unmodeled complexity
(i.e., deviations from self-similarity) in the galaxy population,
rather than by observational errors.

\section{Analysis and Results}

We first simultaneously constrain all four parameters ($\Upsilon_*$,
$x$, $f_{cdm}$, $n$) in our homology model. The best-fit model has
$\chi^2 = 44.1$ for $N_{DOF} = 18$ (with $\delta_{scale} = 1$). The
rms scatter ($0.15$ in $\log M$) is significantly larger than can be
accounted for by the assumed observational uncertainties. Given our
estimates of the measurement errors, the intrinsic scatter is roughly
30\% in mass. Fig.~1 shows the pairwise parameter constraints. The
contours correspond to $\Delta \chi^2 = 1$, $2.30$, $4$ and
$6.17$. Unless otherwise noted, these and all subsequent
$\Delta\chi^2$ values involve the rescaled $\chi^2$, in which
$\delta_{scale}$ is set so that the best-fit model has $\chi^2 =
N_{DOF}$.

There are well-defined regions allowed by the data, but substantial
degeneracy between parameters. This is expected because the
mass-radius relations constrain the combined (luminous plus dark
matter) radial mass distribution. For example, we require a projected
mass of $\sim 3 \times 10^{11} M_{\odot}$ inside $2 R_e$ for the
typical $L_*$ lens galaxy. This can be achieved by placing more mass
in the luminous component, which requires a larger $\Upsilon_*$, or
more mass in the dark matter component, which requires a smaller
$\Upsilon_*$. This degeneracy is reflected in the $\Upsilon_* -
f_{cdm}$ panel of Fig.~1. The mass-radius relations also prefer some
overall shape, or concentration, for the combined profile. To achieve
a given mass concentration, we can place more of the mass in the
stellar component if the CDM slope is shallower, and less if the CDM
slope is steeper. These degeneracies ($\Upsilon_* - n$, $f_{cdm} - n$)
are also clearly seen in Fig.~1.

The slope $x$, which describes the increase in stellar mass-to-light
ratio with luminosity, is uncorrelated with the other three parameters
(Fig.~1). Optimizing over the other parameters, we find $x =
0.14^{+0.16}_{-0.12}$ at $68\%$ confidence ($\Delta \chi^2 < 1$), and
$-0.08 < x < 0.49$ at $95\%$ confidence ($\Delta \chi^2 < 4$).  This
is the first constraint on $x$ from strong lensing.

Our analysis robustly detects the presence of dark matter in the form
of a mass component that is more spatially extended than the
light. Dark matter contributes several tens of percent to the
projected mass inside $2R_e$. This result can be quantified in a
variety of ways, for different assumptions:
\begin{itemize}

\item
A model in which mass traces light ($f_{cdm} = 0$) has $\Delta \chi^2
= 10.7$ with respect to the overall best fit, and is therefore
rejected at $>99\%$ confidence. It is too centrally concentrated to
account for the ensemble of aperture mass-radius relations. We note in
passing that such a model would require $\Upsilon_* \simeq 11
\Upsilon_{\odot}$ in the rest frame $B$ band, only slightly higher
than the value of $\Upsilon_* = (7.8 \pm 2.7) \Upsilon_{\odot}$
determined by Gerhard et al.\ (2001) from dynamical analyses of the
central regions of local (nearly $L_*$) early-type galaxies.  

\item 
Optimizing over the other three parameters, we find $f_{cdm} > 0.36$
at 68\% confidence, and $f_{cdm} > 0.08$ at 95\% confidence.

\item
For a shallow CDM density slope of $n=1$, as suggested by the NFW
profile, the best-fit model has $f_{cdm} = 0.22$, and $\Delta \chi^2 =
2.4$ with respect to the best overall model. The allowed range is
$0.12 < f_{cdm} < 0.32$ at $68\%$ confidence, and $0.08 < f_{cdm} <
0.45$ at $95\%$ confidence.

\item
For a steeper CDM density slope of $n=1.5$, as suggested by the Moore
et al.\ (1999b) profile, the best-fit model has $f_{cdm} = 0.43$, and
$\Delta \chi^2 = 0.8$ with respect to the best overall model. The
allowed range is $0.27 < f_{cdm} < 0.57$ at $68\%$ confidence, and
$0.14 < f_{cdm} < 0.71$ at $95\%$ confidence.

\end{itemize}

Because there are significant degeneracies between parameters, we
cannot place a tight constraint on the slope of the dark matter
component: $1.44 < n < 2.20$ at 68\% confidence, with an upper bound
of $n < 2.33$ at 95\% confidence. However, the favored models have
dark and luminous components which sum to produce very similar mass
profiles over the radial range probed by the lensed images. This is
illustrated in Fig.~2, which shows the projected mass $M(R)$ versus
$R$ for all models with $\Delta \chi^2 < 2.30$ in the $f_{cdm} - n$
plane. We see that the mass profile closely tracks a pure isothermal
($M\propto R$) model.

In the scale-free limit ($\Upsilon_* \rightarrow 0$ and $f_{cdm}
\rightarrow 1$, see \S 4), we find $n=2.07 \pm 0.13$ at 68\%
confidence, consistent with isothermal or slightly steeper
profiles. The 95\% range is $1.83 < n < 2.33$. The best-fit scale-free
model has $\Delta \chi^2 = 0.1$ relative to the best overall model, so
it provides a good approximation to the mass distributions in lens
galaxies.

Fig.~3 demonstrates how the ensemble of mass-radius relations traces
the radial mass profile. The solid line shows the scaled aperture mass
$M(R)/(L/L_*)^{1+x}$ as a function of the scaled aperture radius
$R/R_e$, predicted by the best-fit power-law model. Quad and ring
systems are plotted at a single radius $R_{Ein}$, with enclosed mass
$\Sigma_{cr} \pi R_{Ein}^2$. Doubles are more difficult to depict, as
they constrain a combination of two masses and radii. Note, however,
that any mass profile can be normalized such that the mass-radius
relation for a specific double is satisfied exactly. The resulting values
of $M(R_1)$ and $M(R_2)$ in the context of the best-fit profile are
plotted for each double in Fig.~3. While the locations of the points
obviously depend on the assumed model, the best-fit profile will
minimize their scatter about the model curve. In this way we see that
doubles also help map the mass profile.

The radial coverage offered by the current lens sample is not uniform
(Fig.~3). Quad and ring systems, which provide the strongest mass
constraints, mostly lie in the range $1 \la R_{Ein}/R_e \la 4$. There
are two notable outliers: Q2237+030 (Huchra et al.\ 1985), which
probes the smallest radial scale ($R_{Ein}/R_e \simeq 0.2$), and
MG2016+112 (Lawrence et al.\ 1984), which probes the largest radial
scale ($R_{Ein}/R_e \simeq 7$). Because these lenses constrain the
extremes of the density profile, it is important to address how much
they affect our fits. We find that dropping either of the systems does
not alter the primary conclusions that the total mass profile is
nearly isothermal, or that $x\simeq 0.14$ is favored. Some of the
other constraints are slightly affected, and this can be understood
based on the unique characteristics of each lens. First, MG2016+112
probes a scale much larger than the effective radius, and its aperture
mass will be dominated by dark matter in most models. This system is
therefore important for constraining the properties of the CDM
component. Removing it weakens the exclusion of models in which mass
traces light to about $95\%$ confidence, and significantly weakens the
upper limit on $n$ in the limit $f_{cdm} \rightarrow 0$. Profile
constraints in the scale-free limit ($f_{cdm} \rightarrow 1$),
however, are not significantly changed.  Second, Q2237+030 probes a
scale much smaller than the effective radius, and its aperture mass
will be dominated by luminous matter in most models. Consequently,
this system contributes minimally to constraints on the CDM
component. Dropping it has little effect on any of the model
parameters, even those related to the stellar mass.

Finally, it is interesting to dissect the distribution of scatter
about the best-fit homology model. Despite comprising only 7 of the 22
systems, doubles account for 56\% of the $\chi^2$. The additional
scatter for doubles is greater than we would expect from the
quadrupole smearing, and may indicate some unmodeled effect which we
do not presently understand. Dropping all doubles would significantly
improve the fit ($\chi^2 = 17.1$ for $N_{DOF} = 11$, prior to
rescaling) and tighten the resulting parameter constraints, but would
not alter the major conclusions. For example, a model in which mass
traces light would be rejected more strongly ($\Delta \chi^2 = 19.6$),
and constraints on the density slope in the scale-free limit would be
$n=2.04\pm 0.09$ (68\% C.L.). We also note that a pair of lenses are
particularly strong contributors to the scatter. The two-image system
Q0142--100 (Surdej et al.\ 1987) is the farthest outlier, just as it
is in our FP analysis (Rusin et al.\ 2003), and accounts for almost
$1/4$ of the $\chi^2$. Because the galaxy appears to be much brighter
than expected, it greatly over-predicts aperture masses. The ring
system MG1131+0456 (Hewitt et al.\ 1988) is the second largest
contributor to the $\chi^2$, accounting for $1/7$ of its value,
despite being assigned a rather large fit tolerance due to its
estimated source redshift. The lens galaxy can be brought into better
agreement with the other galaxies if the lensed source is at a
substantially higher redshift than the assumed median of $z_s
=2$. While our sample is too small to reasonably remove these two
outliers, doing so would improve the fit ($\chi^2 = 28.3$ for $N_{DOF}
= 16$, prior to rescaling) but not affect the major conclusions.  For
example, a model in which mass traces light would be rejected somewhat
more strongly ($\Delta \chi^2 = 11.1$), and constraints on the density
slope in the scale-free limit would be $n=2.08\pm 0.10$ (68\% C.L.).

\section{Applications}

\subsection{Time Delays and the Hubble Constant}

The time delays between gravitationally lensed images depend on a
combination of the Hubble constant and the mass distribution of the
lensing galaxy. The Hubble constant inferred from a measured time
delay can be approximated as
\begin{equation}
H_0 = A(1-\langle \kappa \rangle) + B \langle \kappa \rangle
(\eta-1) +C \ ,
\end{equation}
where $\langle \kappa \rangle \equiv \langle \Sigma \rangle /
\Sigma_{cr}$ is the mean scaled surface mass density in the annulus
defined by the radii of the lensed images, $\eta$ is the logarithmic
slope of the density ($\kappa \propto R^{1-\eta}$) within the annulus,
and the constants $A$ and $B$ incorporate the lensing geometry, the
redshifts, and the measured delay (Kochanek 2002). The $A$ term is the
most important, and the $B$ term just contributes a small ($\sim$10\%)
correction.  Kochanek (2002, 2003) investigates ``simple'' time delay
lenses which are dominated by a single galaxy with a precisely
determined centroid relative to the images. Two limiting cases of the
mass distribution are considered: models in which mass traces light
yield $h \simeq 0.7$, while isothermal models yield $h \simeq 0.5$. A
number of recent results from a broader (perhaps less well understood)
sample of lenses fall within the range $0.45 < h <0.65$ (Burud et al.\
2002a, 2002b; Fassnacht et al.\ 2002; Treu \& Koopmans 2002b; Winn et
al.\ 2002) if isothermality is assumed.  In fact, the only lens to
yield $h\simeq 0.7$ for an isothermal model is B0218+357 (Biggs et
al.\ 1999), but the result is currently dominated by systematics
related to the poorly constrained galaxy position. One can therefore
make the case that estimates of the Hubble constant from strong
lensing are in broad agreement, but are systematically lower than the
values favored by both the HST Key Project (Freedman et al.\ 2001) and
the recent WMAP results (Spergel et al.\ 2003). Because Kochanek
(2002, 2003) shows that nearly constant mass-to-light ratio models are
necessary to reconcile time delay measurements with $h\simeq 0.7$,
this would seem to imply that lens galaxies have no extended dark
matter component, in contradiction to virtually every other
observational test.

We can now explore this intriguing problem in the context of our
constraints on self-similar galaxy models. Of the five lenses
considered by Kochanek (2002), three are drawn from our sample. The
two exceptions are B1600+434 (Jackson et al.\ 1995), which has a
late-type lens galaxy (Koopmans, de Bruyn \& Jackson 1998) and is
therefore not described by our models, and RXJ0911+0551 (Bade et al.\
1997), which has a significant cluster perturbation complicating
interpretation of the image radii. Because PG1115+080 (Weymann et al.\
1980) has been explored in detail by Treu \& Koopmans (2002b), we
limit our demonstration to a pair of two-image lenses, SBS1520+530
(Chavushyan et al.\ 1997) and HE2149--2745 (Wisotzki et al.\ 1996).

Because the favored homology models produce very similar mass
profiles, $\langle \kappa \rangle$, $\eta$ and the implied $H_0$ do
not vary much along the $f_{cdm} - n$ degeneracy stripe. These three
quantities are plotted for SBS1520+530 and HE2149--2745 in Figs.~4 and
5, respectively, by normalizing the models to reproduce the
mass-radius relations exactly. The values of $H_0$ use the
coefficients $A$, $B$ and $C$ from Kochanek (2002). The uncertainties
in these coefficients are propagated to $H_0$ using Monte Carlo
procedures. Assuming that each model ($f_{cdm,i}$,$n_j$) has a
likelihood $p_{ij} \propto \exp(-\Delta \chi_{ij}^2/2)$, constraints
on $H_0$ can be determined from a Bayesian analysis. These constraints
are shown in panel (d) of Figs.~4 and 5. For SBS1520+530, $h = 0.58\pm
0.08$ (68\% C.L.). For HE2149--2745, $h = 0.55\pm 0.10$ (68\%
C.L.). The values of $h$ come out slightly higher than those obtained
by Kochanek (2002) for an isothermal profile, because our favored
profiles are slightly steeper than $n=2$. The Hubble constants are
still lower than those favored by the HST Key Project ($h=0.72 \pm
0.08$; Freedman et al.\ 2001) and WMAP data ($h=0.72\pm 0.05$; Spergel
et al.\ 2003), but the discrepancy is only about $1\sigma$ in our
analysis.

\subsection{Flux Ratio Anomalies and Microlensing}

Simple, smooth models for the galaxy mass distribution have great
difficulty reproducing the observed flux ratios in four-image lenses
(e.g., Dalal \& Kochanek 2002; Metcalf \& Zhao 2002; Chiba 2002;
Keeton et al.\ 2003). Moreover, flux ratio measurements in both the
radio and optical appear to violate fundamental symmetry arguments
which predict nearly equal magnifications for merging image
pairs. Such flux ratio ``anomalies'' are almost certainly due to
small-scale structure in the gravitational potential (Kochanek \&
Dalal 2003), which can significantly perturb magnifications (e.g., Mao
\& Schneider 1998) while leaving the image positions virtually
unaltered. There is a debate regarding the identification of the
perturbers. CDM simulations (Moore et al.\ 1999a) predict significant
substructure (with mass scale $M \ga 10^6 M_{\odot}$) in galaxy halos,
and Dalal \& Kochanek (2002) demonstrate that the substructure mass
fraction required to account for radio flux ratio anomalies is in line
with $N$-body results. Stars are another source of substructure in the
gravitational potential (Witt et al.\ 1995), and microlensing-induced
time variability has been unambiguously detected in the light curves
of Q2237+030 (Wozniak et al.\ 2000). Schechter \& Wambsganss (2002)
have pointed out that there is a tendency for saddle point images to
be demagnified compared to our expectations for smooth lens models
(see also Kochanek \& Dalal 2003). They demonstrate that stellar
microlensing produces this effect only if the stars account for a
relatively small fraction ($0.15 \la \kappa_{lum}/\kappa \la 0.30$) of
the surface mass density at the image radius.\footnote{To ward off
possible confusion of terminology, we note that while $f_{cdm}$ is a
global parameter describing the projected dark matter mass fraction
inside $2 R_e$, $\kappa_{lum}/\kappa$ is a local parameter describing
the fraction of surface mass density in the form of stars at the
radius of a lensed image.}  If $\kappa_{lum}/\kappa$ is too high, then
the asymmetric effects on saddles and local minima disappear; if
$\kappa_{lum}/\kappa$ is too low, then the anomalies are too rare. We
can now test whether our two-component mass models fall within the
favored range.

In the context of our homology model, the value of
$\kappa_{lum}/\kappa$ at a given radius $R/R_e$ depends only on the
dark matter abundance parameter $f_{cdm}$ and density slope $n$.
Fig.~6a plots $\kappa_{lum}/\kappa$ at $2R_e$, a typical radial
distance for lensed images. Since we require models with an extended
dark matter component, lower values of $\kappa_{lum}/\kappa$ are
favored at $2 R_e$ due to the fact that the surface brightness of the
galaxy has already decreased substantially at this radius, while the
dark matter density has not. Assuming that each model
($f_{cdm,i}$,$n_j$) has a likelihood $p_{ij} \propto \exp(-\Delta
\chi_{ij}^2/2)$, constraints on $\kappa_{lum}/\kappa$ can be
determined from a Bayesian analysis. These constraints are shown in
Fig.~6b. At $R_e$, just about any value of $\kappa_{lum}/\kappa$ is
permitted. At $2 R_e$, however, the results are quite restrictive,
favoring $\kappa_{lum}/\kappa < 0.31$ (68\% C.L.). Hence, our models
are consistent with the values of $\kappa_{lum}/\kappa$ needed to
produce the relative demagnification of saddle point images observed
in flux ratio anomalies.

\section{Discussion}

We have constrained the typical mass profile of early-type lens
galaxies using an ensemble of aperture mass-radius relations from 22
gravitational lenses.  The different galaxies are combined using a
self-similar mass model consisting of four parameters: the present-day
($z=0$) stellar mass-to-light ratio in the $B$ band for an $L_*$
galaxy ($\Upsilon_*$), its dependence on galaxy luminosity ($x$, where
$\Upsilon \propto L^x$), the projected mass fraction of dark matter
inside 2 effective radii ($f_{cdm}$), and the logarithmic density
slope of the CDM ($n$, where $\rho \propto r^{-n}$). Despite
significant parameter degeneracies, the favored models have dark and
luminous components which sum to produce very similar total mass
profiles over the radial range probed by the lensed images. In the
scale-free limit we find $n=2.07 \pm 0.13$ (68\% C.L.), consistent
with isothermal or slightly steeper profiles. Two-component models
imply that the dark matter mass fraction is not very high: $0.12 <
f_{cdm} < 0.32$ for an $n=1$ halo (NFW 1997), or $0.27 < f_{cdm} <
0.57$ for a steeper $n=1.5$ halo (Moore et al.\ 1999b), both at 68\%
confidence.  Even so, the need for an extended dark matter component
in the inner regions of the galaxies is significant, as a model in
which mass traces light is ruled out at $>99\%$ confidence.  These are
the best statistical constraints yet obtained on the typical radial
mass profile in gravitational lens galaxies, and they add to evidence
that the mass distribution is nearly isothermal on the scale of a few
effective radii.

One might expect lenses to be a biased sample of early-type galaxies,
because deflectors with more centrally-concentrated profiles have
larger lensing cross sections per unit mass. Consequently, if the
galaxy population exhibits a range of profiles, then lensing would
tend to select galaxies with steeper mass distributions. Note,
however, that profile constraints for the lens galaxy population are
consistent with the nearly isothermal profiles favored by dynamical
and X-ray studies of local ellipticals, which are not selected on the
basis of mass concentration. This would appear to strengthen the
argument that early-type galaxies do not exhibit much structural
variety, and therefore, that gravitational lenses are typical members
of this galaxy population. Of course, even if lenses were a biased
galaxy sample, it would have no effect on applications such as Hubble
constant determination, which only requires an understanding of the
mass profiles in lens galaxies.

Our results have implications for the relationship between dark and
luminous matter in early-type galaxies, and can be compared with
recent studies involving the stellar dynamics of local galaxies
(Gerhard et al.\ 2001) and lenses at intermediate redshift (KT), the
number and size distribution of lenses (Keeton 2001), and the
existence of a tight fundamental plane (Borriello, Salucci \& Danese
2003). While the use of different mass models and priors makes a
detailed comparison difficult, we can nonetheless survey the basic
results for consistency.

First, consider evidence for the existence and abundance of dark
matter within a few optical radii.  Our results rule out a galaxy
population in which mass traces light at $>99\%$ confidence, and imply
a small but significant dark matter mass fraction inside $2 R_e$. KT
likewise reject constant mass-to-light ratio models based on the
measured velocity dispersions of lens galaxies. In comparison, Gerhard
et al.\ (2001) claim 10--40\% dark matter within the volume defined by
$R_e$, but caution that some elliptical galaxies show no evidence for
dark matter at this scale. Coming from the other side, Keeton (2001)
uses a variety of lensing statistics to place an upper limit of $40\%$
dark matter within the volume defined by $2 R_e$. Borriello et al.\
(2003) also suggest that a cuspy dark matter halo can contribute
little to the mass inside a few effective radii, if early-type
galaxies are to occupy a tight fundamental plane. We can compare these
measurements to the predictions of the CDM model. Assuming an NFW halo
profile, a cooled baryon fraction of $f_{cool} \equiv
\Omega_{b,cool}/\Omega_M \simeq 0.02$ (from the local census of
baryons by Fukugita, Hogan \& Peebles 1998), and a typical initial
halo concentration ($c \simeq 9$; Bullock et al.\ 2001), the adiabatic
compression models of Keeton (2001) predict $f_{cdm} \simeq
0.6$--$0.8$. The predicted value of $f_{cdm}$ can be reduced slightly
by increasing the cooled baryon fraction, but even maximally efficient
cooling ($f_{cool} = \Omega_b/\Omega_M = 0.17$; Spergel et al.\ 2003)
yields $f_{cdm} \simeq 0.4$--$0.5$. Because the above suite of
measurements seem to imply lower dark matter mass fractions in the
central regions of early-type galaxies, they may support the evidence
from late-type galaxies (e.g., McGaugh \& de Blok 1998; de Blok et
al.\ 2001; Salucci 2001; de Blok \& Bosma 2002) that halos are less
concentrated than predicted by numerical simulations of CDM (e.g.,
Moore et al.\ 1999b; Bullock et al.\ 2001).

Next consider the dependence of mass-to-light ratio on luminosity.  We
find that $\Upsilon \propto L^{0.14^{+0.16}_{-0.12}}$ (68\% C.L.), the
first such constraint from strong lensing. While this result strictly
applies to the stellar mass-to-light ratio, the self-similarity of our
model means that it also applies to the total mass-to-light ratio
inside any fixed fraction of the effective radius. The luminosity
dependence is slightly shallower than the slope of $0.5 \la x \la 0.7$
measured by Gerhard et al.\ (2001) in the $B$ band using stellar
dynamics. Each of these results is broadly consistent with the value
of $x\simeq 0.30\pm0.05$ obtained by interpreting the $B$-band FP
under the assumption of structural homology (e.g., Jorgensen et al.\
1996). Similar investigations have recently been performed at longer
wavelengths.  Bernardi et al.\ (2003) measure $x=0.14\pm 0.02$ in the
$r^*$ band using virial mass-to-light ratios, and derive a similar
value from their FP slopes.  In addition, Borriello et al.\ (2003)
show that scatter in the local $r$-band FP is minimized by $x \simeq
0.2$. Finally, it is interesting to note that the above estimates of
$x$, determined at the scale of the optical radius, are similar to the
relations between virial mass and luminosity determined from weak
lensing and satellite dynamics: Guzik \& Seljak (2002) and Prada et
al.\ (2003) find $M_{vir} \propto L^{1.2-1.6}$, while McKay et al.\
(2001, 2002) favor $M_{vir} \propto L$. The broad consistency of these
scaling laws suggests that the mass distribution in early-type
galaxies is closely related to the light over many orders of magnitude
in radius. 

We have applied our constraints on the mass distribution of early-type
galaxies to a pair of interesting problems. First, we find that the
measured time delays in SBS1520+530 and HE2149--2745 favor a Hubble
constant of $h\simeq 0.55-0.60$. This value is slightly higher than
that derived by Kochanek (2002) using isothermal models, since we
prefer mass profiles that are slightly steeper than $n=2$. The derived
$H_0$ is still systematically lower than recent results (Freedman et
al.\ 2001; Spergel et al.\ 2003), although the discrepancy is only
about $1\sigma$ in our analysis, as the statistical uncertainties in
the mass profile are still significant. Second, we have estimated the
fraction of the surface mass density in the form of stars
($\kappa_{lum}/\kappa$) at the radii of lensed images. For images at
$2R_e$, our models favor $\kappa_{lum}/\kappa < 0.31$ (68\%
C.L.). This is consistent with the values at which stellar
microlensing can reproduce the observed flux ratio anomalies, in
particular the tendency for saddle-point images to be more strongly
perturbed than minima (Schechter \& Wambsganss 2002).

Much more work is needed to fully understand the structure of
early-type galaxies, and the implications for galaxy formation
theories. Because of its unique advantages, gravitational lensing is
certain to contribute significantly to this goal. Improved statistical
tests to constrain the mass profiles of early-type galaxies require
larger samples of lenses. Particularly important is the addition of
new quad and ring systems which probe the mass distribution on small
($R_{Ein} < R_e$) and large ($R_{Ein} > 4 R_e$) scales, and therefore
fill in the tails of our radial coverage. As always, the key to
turning recently discovered systems into useful astrophysical tools is
the acquisition of high-quality HST photometry and ground-based
spectroscopy. Expanded samples could allow us to replace our
simplistic power-law model for the dark matter halo with more
realistic profiles, and consider structural dependencies on luminosity
or color. In this way we may be able to properly investigate sources
of scatter in our homology model, and perhaps quantify or constrain
the structural diversity of the early-type galaxy population.  With
regard to lensing determinations of the Hubble constant, it is vital
to increase the number of lenses that have both direct profile
constraints and well-determined time delays. Currently there is only one
such system (PG1115+080; Schechter et al.\ 1997; Treu \& Koopmans
2002b).\footnote{A number of profile constraints have been claimed for
the time delay lens Q0957+561 (e.g., Grogin \& Narayan 1996), but
Keeton et al.\ (2000) demonstrates that existing models fail to
properly reproduce lensed extended emission from the quasar host
galaxy.} Programs are now underway to both measure more time delays,
and obtain profile constraints on current systems by measuring stellar
velocity dispersions or mapping extended emission from the lensed host
galaxies. Such observations should greatly illuminate current
discrepancies related to time delays and the Hubble constant, and,
more generally, improve our understanding of the mass distribution in
early-type galaxies.

\acknowledgements

We thank Josh Winn and the anonymous referee for offering comments and
suggestions which greatly improved this manuscript. We acknowledge the
support of HST grants GO-7495, 7887, 8175, 8804, and 9133. We
acknowledge the support of the Smithsonian Institution. CSK is
supported by NASA ATP Grant NAG5-9265.

\clearpage
\begin{deluxetable}{rrcccc}
\tablecaption{Structural Parameters for Lens Systems }
\tablewidth{0pt}
\tablehead{Lens & $z_d$ & $z_s$ & Morph. & Radii ($''$)\\
}
\startdata
    0047--2808  & 0.49 & 3.60  & Quad   & $R_{Ein} = 1.35$ \\
Q0142--100	& 0.49 & 2.72  & Double & $R_1 = 0.38$, $R_2 = 1.86$ \\
   MG0414+0534  & 0.96 & 2.64  & Quad   & $R_{Ein} = 1.19$ \\ 
     B0712+472  & 0.41 & 1.34  & Quad   & $R_{Ein} = 0.71$ \\
   HS0818+1227  & 0.39 & 3.12  & Double & $R_1 = 0.61$, $R_2 = 2.22$ \\
     B1030+074  & 0.60 & 1.54  & Double & $R_1 = 0.18$, $R_2 = 1.39$ \\
  HE1104--1805  & 0.73 & 2.32  & Double & $R_1 = 1.10$, $R_2 = 2.09$ \\
    PG1115+080  & 0.31 & 1.72  & Quad   & $R_{Ein} = 1.15$ \\
   MG1131+0456  & 0.84 & --    & Ring   & $R_{Ein} = 1.05$ \\
 HST14113+5211  & 0.46 & 2.81  & Quad   & $R_{Ein} = 0.86$ \\
 HST14176+5226  & 0.81 & 3.40  & Quad   & $R_{Ein} = 1.42$ \\
     B1422+231  & 0.34 & 3.62  & Quad   & $R_{Ein} = 0.78$ \\
   SBS1520+530  & 0.72 & 1.86  & Double & $R_1 = 0.39$, $R_2 = 1.21$ \\
   MG1549+3047  & 0.11 & 1.17  & Ring   & $R_{Ein} = 1.15$ \\
     B1608+656  & 0.63 & 1.39  & Quad   & $R_{Ein} = 1.14$ \\
   MG1654+1346  & 0.25 & 1.74  & Ring   & $R_{Ein} = 1.05$ \\
     B1938+666	& 0.88 &  --   & Ring   & $R_{Ein} = 0.50$ \\
    MG2016+112  & 1.00 & 3.27  & Quad   & $R_{Ein} = 1.63$ \\
     B2045+265  & 0.87 &  --   & Quad   & $R_{Ein} = 1.14$ \\
  HE2149--2745  & 0.50 & 2.03  & Double & $R_1 = 0.34$, $R_2 = 1.35$ \\
     Q2237+030  & 0.04 & 1.69  & Quad   & $R_{Ein} = 0.88$ \\
     B2319+051  & 0.62 & --    & Double & $R_1 = 0.61$, $R_2 = 0.82$ \\
\enddata 

\tablecomments{Listed for each lens are the galaxy ($z_d$) and source
($z_s$) redshifts, morphology, and relevant radii (Einstein radius
$R_{Ein}$ for quads and rings; image radii $R_1$ and $R_2$ for
doubles).  Uncertainties on these radii are negligible. The image
radii for B2319+051 are the estimated values from Rusin et al.\
(2001). }
\end{deluxetable}

\begin{figure*}
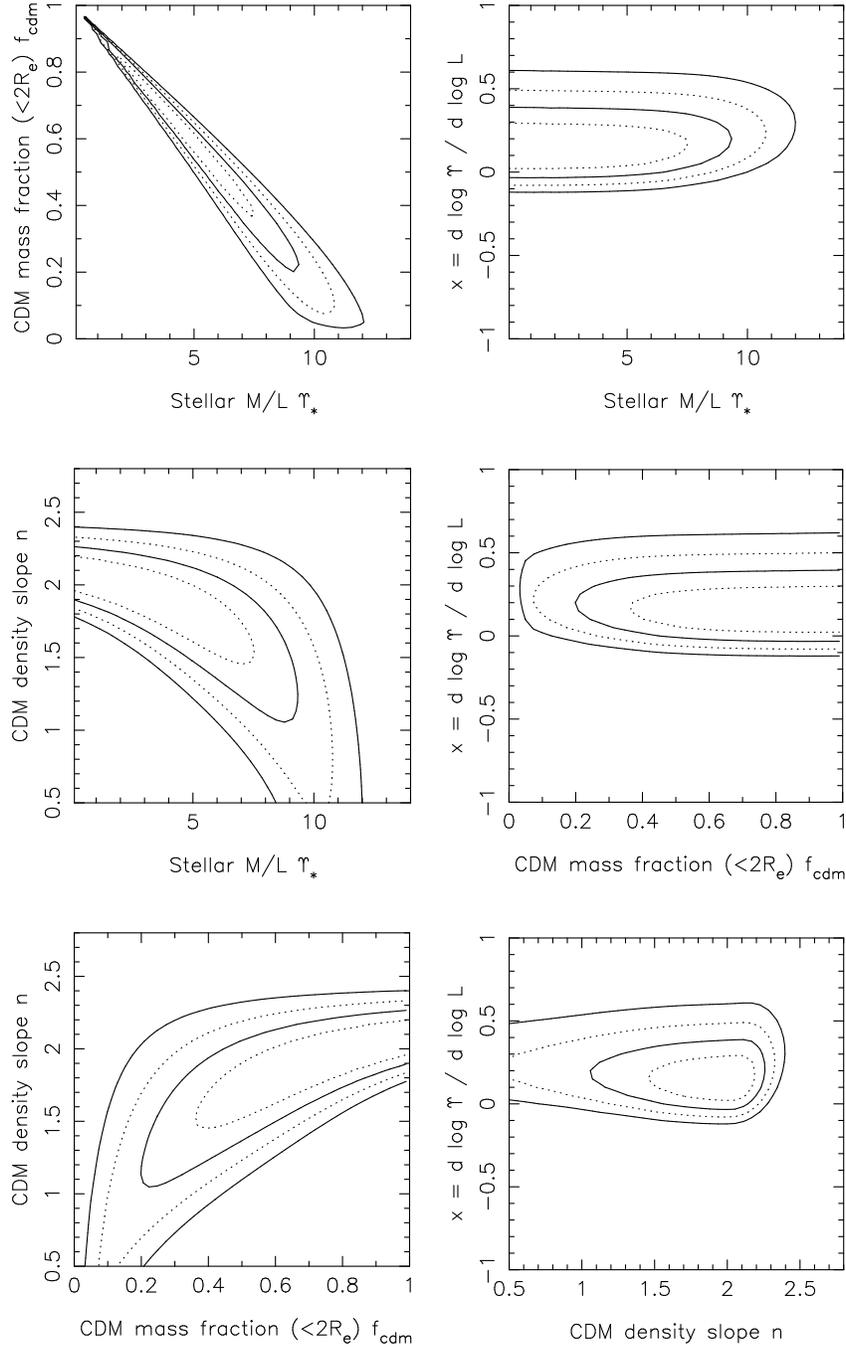

\begin{center}
\begin{tabular}{cc}
\psfig{file=cdm8big_12.ps,width=2.1in}&
\psfig{file=cdm8big_14.ps,width=2.1in}\\
 & \\
\psfig{file=cdm8big_13.ps,width=2.1in}&
\psfig{file=cdm8big_24.ps,width=2.1in}\\
 & \\
\psfig{file=cdm8big_23.ps,width=2.1in}&
\psfig{file=cdm8big_34.ps,width=2.1in}\\
\end{tabular}
\end{center}
\figurenum{1}
\caption{Constraints on the self-similar model. We show pairwise
constraints on the present-day ($z=0$) stellar mass-to-light ratio in
the $B$ band for an $L_*$ galaxy ($\Upsilon_*$, in solar units), its
logarithmic dependence on luminosity ($x$), the projected CDM mass
fraction inside 2 effective radii ($f_{cdm}$), and the logarithmic CDM
density slope ($n$). Solid contours represent $\Delta \chi^2 = 2.30$
and $6.17$, the 68\% and 95\% confidence levels for two parameters.
Dotted lines represent $\Delta \chi^2 = 1$ and $4$, the 68\% and 95\%
confidence levels for one parameter. The errors have been rescaled so
that the best-fit model has $\chi^2 = N_{DOF}$.}
\end{figure*}

\begin{figure*}
\begin{tabular}{c} 
\psfig{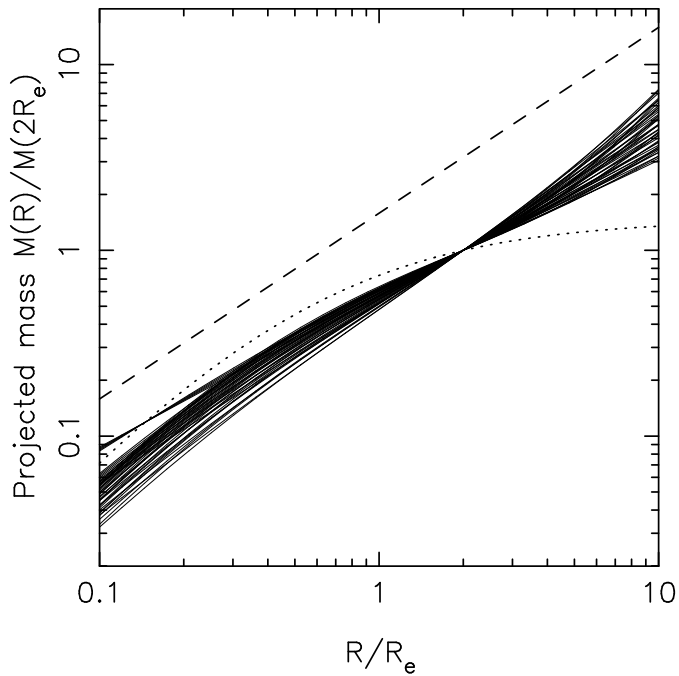}\\
\end{tabular}
\figurenum{2} 
\caption{Mass profiles for allowed two-component homology models. Each
model falls within the $68\%$ ($\Delta \chi^2 < 2.30$) confidence
region of the $f_{cdm}-n$ plane (Fig.~1). Solid lines are the
projected masses inside $R/R_e$, where $R_e$ is the effective
radius. Profiles are normalized to a fixed projected mass at
$R=2R_e$. For comparison we show the de Vaucouleurs profile (dotted
line), and an offset isothermal profile (dashed line). While the
allowed models exhibit a wide range of dark matter abundances, they
all have total mass profiles which are approximately isothermal over
the radial range spanned by the lensed images.}
\end{figure*} 

\begin{figure*}
\begin{tabular}{c} 
\psfig{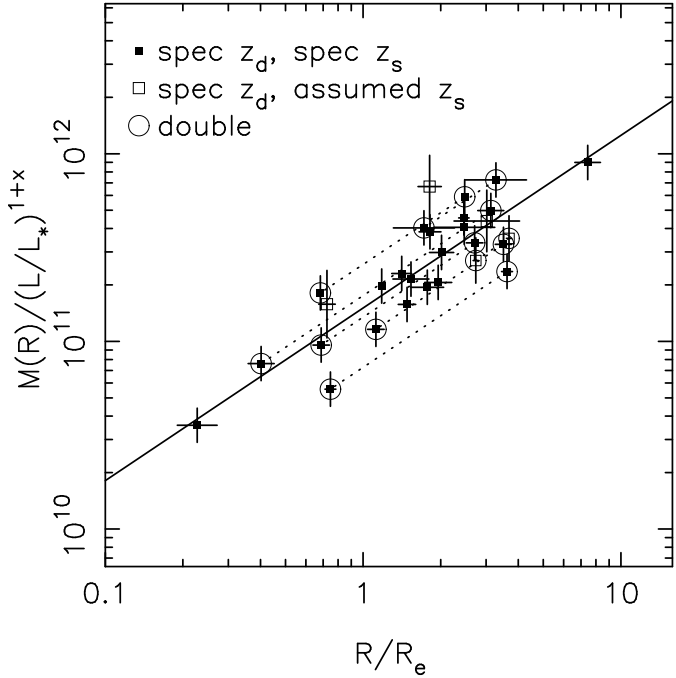}\\
\end{tabular}
\figurenum{3}
\caption{The aperture mass profile in the power-law limit. We plot the
scaled projected mass $M(R)/(L/L_*)^{1+x}$ as a function of the scaled
radius $R/R_e$. The solid line is the prediction of the best-fit
power-law model with $n=2.07$ and $x=0.14$. Data points represent
individual lens systems. Solid squares are lenses with measured source
redshifts; open squares are lenses with estimated ($z_s = 2.0 \pm
1.0$) source redshifts. For quads and rings, a single point is plotted
showing $R_{Ein}$ and mass $\Sigma_{cr} \pi R_{Ein}^2$. For doubles
(big circles), points are plotted at each image radius ($R_1$ and
$R_2$) and connected with a dotted line, and the masses are the
$M(R_1)$ and $M(R_2)$ which would satisfy the mass-radius relation
exactly in the context of this profile (see \S 5).}
\end{figure*}

\begin{figure*}
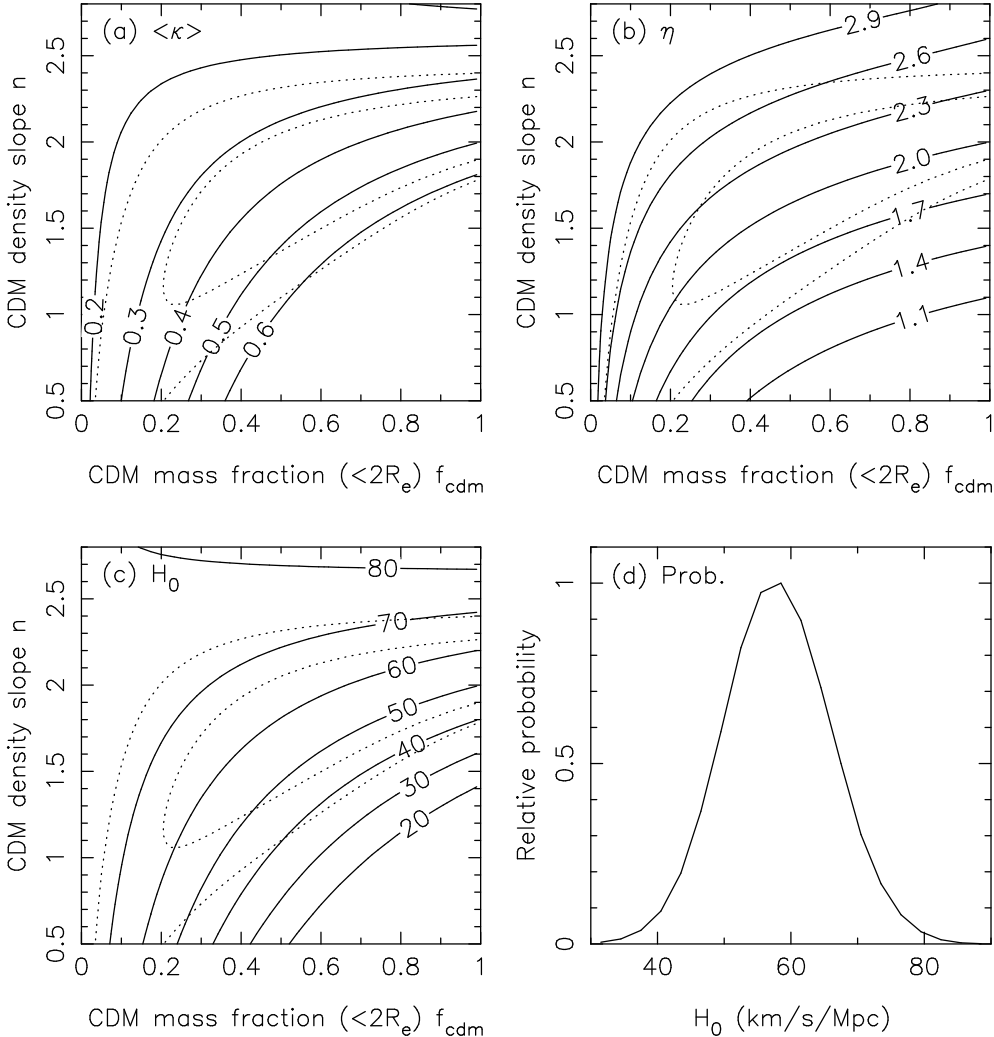

\begin{tabular}{cc} 
\psfig{file=apkap3.1520.1.ps,width=2.5in}&
\psfig{file=apkap3.1520.2.ps,width=2.5in}\\
 & \\
\psfig{file=apkap3.1520.3.ps,width=2.5in}&
\psfig{file=apkap3.1520.ps,width=2.5in}\\ 
\end{tabular}
\figurenum{4} 
\caption{Interpreting the time delays of SBS1520+530. The first three
panels show various quantities as a function of CDM mass fraction
$f_{cdm}$ and CDM density slope $n$. Solid contours indicate (a) the
mean scaled surface density $\langle \kappa \rangle$ inside the image
annulus, (b) the effective density slope ($\eta$) inside the image
annulus, and (c) the derived Hubble constant $H_0$. The models are
normalized to satisfy the mass-radius relation exactly. Note that the
quantities exhibit the expected behavior: $\eta \rightarrow n$ for
$f_{cdm} \rightarrow 1$ and $\langle \kappa \rangle = 0.5$ for a
scale-free isothermal model. Also, we see that an isothermal profile
yields $H_0 \simeq 50\kmsm$, as demonstrated by Kochanek (2002).  The
dotted contours show the 68\% ($\Delta \chi^2 < 2.30$) and 95\%
($\Delta \chi^2 < 6.17$) confidence regions from the homology model
(Fig.~1). Plotted in panel (d) is the relative probability of $H_0$
based on the information displayed in (c). A Hubble constant of $H_0=
(58\pm 8) \kmsm$ is favored (68\% C.L.).  }
\end{figure*}

\begin{figure*}
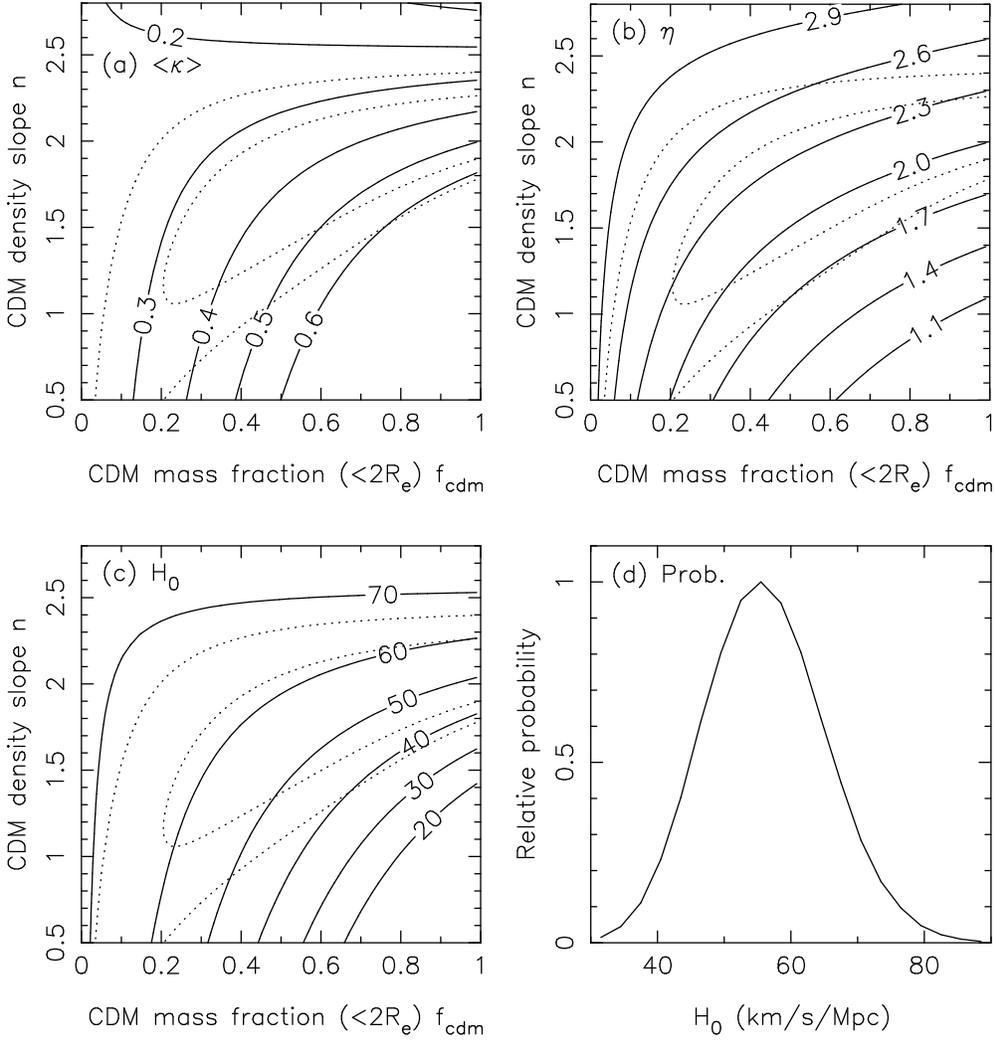

\begin{tabular}{cc} 
\psfig{file=apkap3.2149.1.ps,width=2.5in}&
\psfig{file=apkap3.2149.2.ps,width=2.5in}\\
 & \\
\psfig{file=apkap3.2149.3.ps,width=2.5in}&
\psfig{file=apkap3.2149.ps,width=2.5in}\\ 
\end{tabular}
\figurenum{5} 
\caption{Interpreting the time delays of HE2149--2745. The plots are
analogous to those in Fig.~4. A Hubble constant of $H_0= (55\pm 10)
\kmsm$ is favored (68\% C.L.). }
\end{figure*}

\begin{figure*}
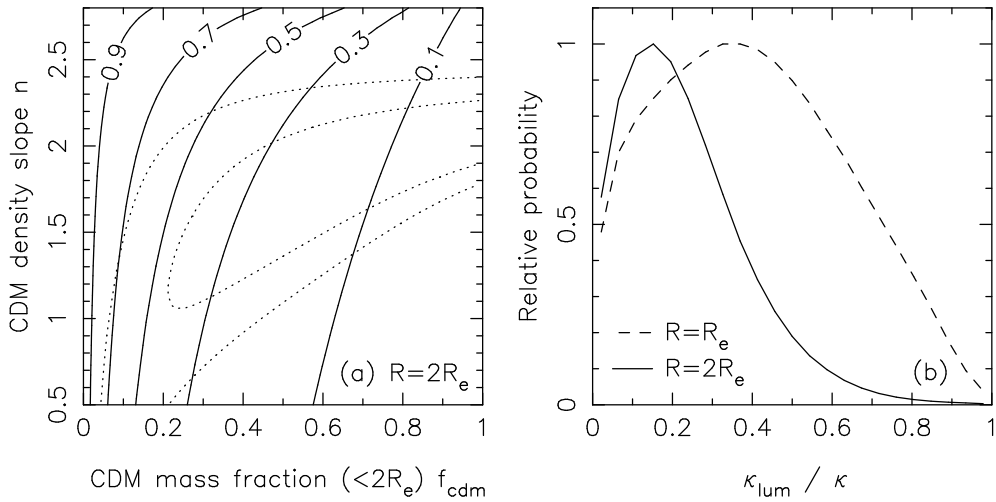

\begin{tabular}{cc} 
\psfig{file=apkap5.generic.re2.2.ps,width=2.5in}&
\psfig{file=apkap5.generic.hist.ps,width=2.5in}\\
\end{tabular}
\figurenum{6} 
\caption{The fraction of the surface mass density in the form of
stars. (a) Plotted are contours of $\kappa_{lum}/\kappa$ at $R=2R_e$
for models in the $f_{cdm}-n$ plane. The dotted contours show the 68\%
($\Delta \chi^2 < 2.30$) and 95\% ($\Delta \chi^2 < 6.17$) confidence
regions from the homology model (Fig.~1). (b) The relative
probability of $\kappa_{lum}/\kappa$ based on the information
displayed in (a). The solid line is the fraction at $R=2R_e$. The
dashed line is the fraction at $R=R_e$.}
\end{figure*}

\end{document}